\documentclass[12pt]{article}
\input epsf
\newcommand{\be}{\begin{equation}}
\newcommand{\ee}{\end{equation}}
\newcommand{\bea}{\begin{eqnarray}}
\newcommand{\eea}{\end{eqnarray}}
\newcommand{\ep}{\varepsilon}

\newcommand{\epse}{\varepsilon^{\prime}/\varepsilon}
\newcommand{\Vckm}{V_{CKM}}
\newcommand{\Vcb}{\vert V_{cb}\vert }
\newcommand{\Vub}{\vert V_{ub}\vert}
\newcommand{\DMd}{\Delta M_{B_d}}
\newcommand{\DMs}{\Delta M_{B_s}}
\newcommand{\asz}{\alpha_s(M_Z)}
\newcommand{\MSb}{\overline{MS}}
\newcommand{\mtms}{m_t^{\overline{MS}}(m_t^{\overline{MS}})}
\newcommand{\mbms}{m_b^{\overline{MS}}(m_b^{\overline{MS}})}
\newcommand{\mcms}{m_c^{\overline{MS}}(m_c^{\overline{MS}})}
\newcommand{\msms}{m_s^{\overline{MS}}(\mu)}
\newcommand{\mdms}{m_d^{\overline{MS}}(\mu)}
\newcommand{\nn}{\nonumber}
\newcommand{\prd}{Phys.~Rev. }
\newcommand{\pl}{Phys.~Lett. }
\newcommand{\np}{Nucl.~Phys. }
\renewcommand{\arraystretch}{1.3}
\begin{document}
\pagestyle{empty} 
\begin{flushright}
BUHEP-99-24\\
RM3-TH/99-9\\
ROME 99/1267
\end{flushright}
\vskip 1cm
\centerline{\Large{\bf{Combined analysis of the unitarity triangle and}}}
\centerline{\Large{\bf{CP violation in the Standard Model}}}
\vskip 1cm
\centerline{\bf{M. Ciuchini$^{a}$, E. Franco$^{b}$, L. Giusti$^{c}$,
V. Lubicz$^{a}$, G. Martinelli$^{b}$}}
\vskip 0.3cm
{\small
\centerline{$^a$ Dipartimento di Fisica, Universit\`a di Roma Tre
and INFN, Sezione}
\centerline{di Roma III, Via della Vasca Navale 84, I-00146 Rome, Italy}
\centerline{$^b$ Dipartimento di Fisica, Universit\`a di Roma ``La Sapienza''
and INFN,}
\centerline{Sezione di Roma, P.le A. Moro 2, I-00185 Rome, Italy}
\centerline{$^c$ Department of Physics, Boston University, Boston, MA 02215 
USA.}
}
\begin{abstract}
We perform a combined analysis of the unitarity triangle and of the CP violating parameter $\epse$
using the most recent determination of the relevant experimental data and, whenever possible,
hadronic matrix elements from lattice QCD.
We discuss the r\^ole of the main non-perturbative parameters and make a
comparison with other recent analyses.
We use lattice results for the matrix element of $Q_8$ obtained without reference to
the strange quark mass.
Since a reliable lattice determination of the matrix element of $Q_6$
is still missing, the theoretical predictions for $\epse$ suffer from large uncertainties.
By evaluating this matrix
element with the vacuum-saturation approximation, we typically find as central value
$\epse =(4\div 7)\times 10^{-4}$.
We conclude that the experimental data suggest large deviation of the value of the matrix element
of $Q_6$ from the vacuum-saturation approximation, possibly due to penguin contractions.
\end{abstract}
\vfill
\begin{flushleft}
October 1999
\end{flushleft}
\eject
\setcounter{page}{1}
\pagestyle{plain}
\section{Introduction}
\protect\label{sec:intro}
The latest measurements of $\epse$
\be
\begin{array}{lll}
\mbox{Re}(\epse)=(28.0 \pm 4.1)\times 10^{-4} & \mbox{KTeV} & \cite{KTeVK99},\\
\mbox{Re}(\epse)=(18.5 \pm 7.3)\times 10^{-4} & \mbox{NA48} & \cite{NA48K99},
\end{array}
\ee
confirm the large value found by NA31~\cite{NA31} and rise the world average to
$\mbox{Re}(\epse)_{\mbox{(WA)}}=(21.2 \pm 4.6)\times 10^{-4}$~\cite{NA48K99}.
Motivated by these results, we present a new study of the unitarity triangle and of CP violation in kaon
decays within the Standard
Model. Our results have been obtained from  a Next-to-Leading Order
(NLO) calculation of $\epse$ combined
with  the constraints  on the Cabibbo-Kobayashi-Maskawa matrix ($\Vckm$) derived from  
measurements of $\Vcb$, $\Vub$, $\ep$, $\DMd$ and the limits on $\DMs$.
This work is an upgraded and improved version of previous studies made by the Rome
group~\cite{ciuc1}--\cite{ciuc3}.  Similar analyses can be found in
the recent literature~\cite{buras1}--\cite{russi}. For previous 
estimates of $\epse$ with a heavy top mass see also \cite{flynn}--\cite{pass}. 
\par
Several features characterize this work:
\begin{itemize}
\item We analyze the constraints on $\Vckm$ together with $\epse$, fully taking 
into account correlation effects. This should be compared with the analyses 
of refs.~\cite{mele,stocchi}, where only the $\Vckm$ constraints were
considered, or with the analyses 
of refs.~\cite{bert1,bert2} and  \cite{paschoslast,pass},
where the input values and errors of the $\Vckm$ parameters used in the study of $\epse$ 
were taken  elsewhere. With respect to the recent study of ref.~\cite{silvestrini}, we present
the results of the analysis of the unitarity triangle 
together with the predictions for $\epse$.
\item In our NLO analysis, we include
the full, correlated  dependence of the coefficients of 
the effective Hamiltonians ${\cal H}^{\Delta
S=2}$, 
${\cal H}^{\Delta B=2}$~ and ${\cal H}^{\Delta 
S=1}$, computed in refs.~\cite{alta}--\cite{urban},
on the relevant parameters, such as $\asz$ or the $\MSb$ top mass, $\mtms$. 
\item We carefully account for the renormalization-scheme dependence of the matrix elements
of the renormalized operators and of the compensating effects in the corresponding Wilson coefficients.
This is specially important, given the large differences in the values of matrix elements of operators
defined with different prescriptions, in particular for $Q_6$ and 
$Q_8$~\footnote{ For a definition
of the operators of ${\cal H}^{\Delta S=1}$, see for example
refs.~\cite{bur3,noi}.}.
\item We address the issue of the  dependence of  theoretical predictions
for $\epse$ on the strange quark mass $m_s$, induced by the standard definition of the $B$ parameters.
In addition, we present results obtained by using the matrix elements of
the electropenguin operators,
$Q_7$ and $Q_8$, computed without any reference to the quark
masses~\cite{giusti}.
\end{itemize}
Our main results are the following. For the unitarity triangle, using experimental
informations on $\Vcb$, $\Vub$, $\ep$, $\DMd$ and $\DMs$, we find
\be
\begin{array}{l l}
\overline\rho=0.16^{+0.08}_{-0.11}\, , &  \overline\eta=0.38^{+0.06}_{-0.05}\, ,\\
\sin(2\alpha)=0.11^{+0.40}_{-0.35}\, , & \sin(2\beta)=0.75^{+0.09}_{-0.09}\, , \\
\gamma=(68^{+15}_{-10})^o\, , & Im\lambda_t=(1.34^{+0.16}_{-0.15})\times 10^{-4}\, .
\end{array}
\label{eq:triangle}
\ee
in good agreement with the results of other recent studies~\cite{mele,stocchi}. A detailed discussion
of this analysis can be found in sect.~\ref{sec:triangle}.

Concerning $\epse$, the major uncertainty still affecting the theoretical predictions
is the lack of a quantitative determination of the matrix element of $Q_6$.
In particular, the status of  lattice
calculations for this matrix element is more confused now than a few years ago. The value
obtained  using staggered fermions,  indicating small deviations from the vacuum-saturation
approximation (VSA)
for the bare operator~\cite{kilcup},  has been found plagued
by huge perturbative corrections~\cite{stagpert} which make it unreliable.
Very recent results
using domain-wall fermions~\cite{newsoni}, on the contrary, correspond to a dramatic violation
of the VSA and predict a sign of the matrix element different from any other non-perturbative approach.
Since this result has been obtained  with a lattice formulation
for which numerical studies began very recently, we think that further scrutiny and
confirmation from other calculations are needed  before using it in phenomenological analyses.

With this {\it caveat} in mind, we prefer to give  firstly our result for $\epse$ in the form
\be
\epse = \left[ (-21.7^{+3.9}_{-4.3})\, \mbox{GeV}^{-3}\times
\langle \pi\pi\vert Q_6^{HV}\vert K\rangle_{I=0}- (6.0^{+1.5}_{-1.8}
)\right]  \times 10^{-4} \, ,
\label{eq:epseq6}
\ee
where the operator $Q_6$ is renormalized at $\mu=2$ GeV in the 't~Hooft-Veltman ($HV$)
renormalization scheme of ref.~\cite{noi} and
the errors are evaluated by varying all the experimental and theoretical
parameters, but $\langle \pi\pi\vert Q_6^{HV}\vert K\rangle_{I=0}$, as explained in
sect.~\ref{sec:method}.
One may use this formula to estimate $\epse$  with any non-perturbative method
able to control the renormalization scale and scheme dependence of $Q_6$ at the NLO.

In the absence of  definite results from the lattice, we take the central value of
$\langle \pi\pi\vert Q_6^{HV}\vert K\rangle_{I=0}$ from the VSA and allow a large variation
of the matrix element using a relative error of
$100\%$. Note that there is an ambiguity in taking this value, since the renormalization
scheme and scale of the operator is unknown in the VSA. By assuming the VSA for the central value
in the $HV$ scheme at $\mu=2$ GeV, we find
\be
\begin{array}{l l}
\epse = (3.6^{+6.7}_{-6.3} \pm 0.5)\times 10^{-4} & \mbox{(Monte Carlo),} \\
-11 \times 10^{-4}  \le \epse \le  27 \times 10^{-4}&    \mbox{(Scanning),}
\end{array}
\label{eq:epse1}
\ee
where the first error comes from the uncertainties on the input parameters and the
second one accounts for the residual renormalization scheme dependence due to
higher orders in the perturbative expansion.
Details on the definition of the errors and the scanning procedure can be found in
sect.~\ref{sec:method}.
Taking the value of the VSA in the NDR scheme, we find instead
\be
\begin{array}{l l}
\epse = (6.7^{+9.2}_{-8.5} \pm 0.4)\times 10^{-4} & \mbox{(Monte Carlo),} \\
-10 \times 10^{-4}  \le \epse \le  30 \times 10^{-4}&    \mbox{(Scanning).}
\end{array}
\label{eq:epse2}
\ee
Note that the difference between eqs.~(\ref{eq:epse1}) and (\ref{eq:epse2})
is not due to the scheme dependence, but to the change in the value of the matrix
element of $Q_6$. For a more detailed discussion of this point, see sect.~\ref{sec:epse}.

We conclude that, with a central value of $\langle\pi\pi\vert Q_6
\vert K\rangle_0$ close to the VSA one, even with a large
error, it is difficult to reproduce the experimental value of $\epse$, for which
a conspiracy of several inputs pushing $\epse$ in the same direction is necessary.
In our opinion, the important message arriving from the  experimental
results is that {\it penguin contractions (eye diagrams), neglected in the VSA,
give contributions to the matrix elements definitely larger than
their factorized  values.} This interpretation provides a unique dynamical mechanism to
account for both the  $\Delta I=1/2$ rule and a large  value of $\epse$ within the Standard Model,
whereas other arguments, as those
based on a low value of  $\msms$, would leave the $\Delta I=1/2$ rule unexplained.

This paper is organized as follows. In sect.~\ref{sec:method}, we describe the methods used
in the phenomenological analysis. Our study of the unitarity triangle can be found
in sect.~\ref{sec:triangle}. Details on the calculation of $\epse$ are given in sect.~\ref{sec:epse}.
In both cases, we make a comparison with other recent analyses.
Section~\ref{sec:conclusion} contains our conclusions.

\section{Analysis method}
\label{sec:method}
The analysis of the unitarity triangle is based on  a comparison of theoretical
expressions for $\ep$, $\DMd$ and $\DMs$ with the
measurements/bounds and on the experimental determination
of $\Vub$ and $\Vcb$. Several parameters enter the theoretical expressions of the above
quantities.  They can be classified in two groups: ``experimental'' quantities,
such as the top and $W$ masses, $\alpha_s(M_z)$, etc., and theoretical ones, such as the Wilson
coefficients, which are computed at the NLO in perturbation theory, and the hadronic matrix elements.

We now describe the procedure used in our combined analysis of $\epse$ and the
unitarity triangle.
\begin{enumerate}
\item For all the parameters
which determine $\ep$, $\DMd$ and $\DMs$, we extract randomly the experimental quantities
with gaussian
distributions and the theoretical ones with flat distributions~\footnote{
In the latter case,
when  for a generic variable $x$  we give the average and the error,
$\bar x \pm dx$,
this means that we extract $x$ with flat probability between
$\bar x-dx$ and $\bar x +dx$.}.  The latter include, for example, $\hat B_K$, the $b$-quark
mass, which enters as a threshold in the evolution of the Wilson
coefficients, etc.
We have some remarks to make on the choice of the error distributions.
In many cases, the main
systematic  error in the determination of the experimental quantities,
such as $\vert V_{ub}\vert$,
comes from the theoretical uncertainties. One may argue that
systematic errors coming from the theory
(and experimental-systematic errors as well) should correspond
to flat distributions. In practice, it may be difficult to
disentangle statistical and systematic errors affecting some
input parameters.
However, as discussed below, the actual choice of the error distributions has
a rather small influence on the final results.
For these reasons, we have assumed for  all the experimental
quantities gaussian distributions, with a width obtained by combining in
quadrature all errors.

\item Among the quantities which are  extracted with a gaussian
distribution, there are also $\Vub$ and $\Vcb$; for a given set of
extracted values
of $\Vub$, $\Vcb$ and $\lambda=\vert V_{us}\vert$, we determine $\sigma=
\sqrt{\rho^2+\eta^2}= \Vub / (\lambda \Vcb)$.
\item For a given value of $\sigma$ (and of all the other relevant parameters),
we extract  a value of $\ep$ with gaussian distribution, and find the solutions of the equation
\bea
\ep=\ep^{th}(\mtms,\alpha_s(M_z),\dots, \hat B_K,\sigma,\delta) \,
\label{eq:ep}
\eea
with respect to $\delta$, which is the CP violation phase in the
standard parameterization of $V_{CKM}$ as adopted by the PDG~\cite{PDG}, with $0 \le \delta \le \pi$.
In eq.~(\ref{eq:ep}),
$\ep^{th}(\mtms,\alpha_s(M_z),\dots, \hat B_K,\sigma,\delta)$ is the theoretical
value computed for that  given set of random parameters and $\ep$ is the extracted value.
The explicit expression
of $\ep^{th}$ can be found for example in eq.~(10) of ref.~\cite{ciuc2}.
In general one finds
two independent solutions for $\delta$. For any set of extractions,
 this fixes two  independent
sets of values for $\rho=\sigma \cos \delta$ and $\eta=\sigma \sin \delta$.
In the following, the set of all the extracted parameters and of one of the
two solutions for $\rho$ and $\eta$ will be denoted as
{\it event}~\footnote{
Thus for any set of random variables we have in general two {\it events}. It
may happen that there is no solution, in this case the {\it event} is disregarded.}.
When we  consider also $\epse$, {\it event}  denotes all the random
variables,
including the matrix elements (and further parameters) which enter the
calculation of this quantity.
\item For a given {\it event},  $n_i$,  we compute a statistical weight
defined as
\bea && {\cal W}_i=\exp \Bigg[
-\frac{1}{2}\left(\frac{\DMd^{exp}-\DMd^{th}(n_i)}{d\DMd}\right)^2 \nn\\
&&-\frac{1}{2} \left(\frac{a[\DMs^{th}(n_i)] -1}
{da[\DMs^{th}(n_i)]}\right)^2 \Bigg] \times
J^{-1}\left[\left(\vert  V_{ub}\vert,\ep^{th}\right),\left(\rho,\eta
\right)\right]
\label{eq:sw} \, , \eea
where  $\DMd^{th}(n_i)$ and $\DMs^{th}(n_i)$ are  the theoretical values of
$\DMd$ and $\DMs$ computed with the {\it event} $n_i$;  $a[\DMs]$ and
$da[\DMs]$
are the average value and error of the oscillation amplitude for $B^0_s$--$\bar B^0_s$
mixing, introduced in  ref.~\cite{moser}~\footnote{ The values of $\bar a$ and
$da$ in bins of $\Delta M_{B_s}$  are produced by the LEP ``B Oscillation Working Group".
We thank A. Stocchi for providing us with these numbers.}.
$J\left[\left(\vert V_{ub}\vert,\ep^{th}\right),\left(\rho,\eta
\right)\right]$ is the Jacobian relating $\vert V_{ub}\vert$ and
$\ep^{th}$ to $\rho$ and $\eta$. With this weight factor, our procedure
coincides with the method followed in ref.~\cite{stocchi}: in that
case they extract with flat distributions $\rho$ and $\eta$, compute
$\vert V_{ub}\vert(\rho,\eta) $ and $\ep(\rho,\eta)$ and include
in  the statistical weight a factor
\be \exp \left[
 -\frac{1}{2}\left(\frac{\vert V_{ub}^{exp}\vert-\vert V_{ub}\vert(\rho,\eta) }
 {d\vert V_{ub}\vert}\right)^2
 -\frac{1}{2}\left(\frac{\ep^{exp}-\ep(\rho,\eta)}{d\ep}\right)^2\right] \, , \ee
 instead of $J^{-1}$.
Since the error on the measurement of $\ep$ is tiny,   most of the
extractions of $\rho$ and $\eta$ in the interval $[0,1]$, correspond
to very small  statistical weights. Thus, the method of
ref.~\cite{stocchi} demands a large number of extractions in order
to obtain a significative statistical sample. Our procedure is, in this
respect, more efficient.
\renewcommand{\arraystretch}{1.0}
\begin{table}[t]
\begin{center}
\begin{tabular}{|c|c|}
\hline
Constant & Values \\ \hline
$G_F$ & $1.16639\times 10^{-5}\, \mbox{GeV}^{-2}$ \\
$\alpha_{e}(M_z)$ & $7.8125\times 10^{-3}$\\
$\sin^2\theta_w$ & 0.23154 \\
$f_{\pi}$ & 0.1307 GeV \\
$f_K$ & 0.1598 GeV \\
\hline
$M_W$ & 80.41 GeV \\
$M_Z$ & 91.1867 GeV \\
$M_{B_d}$ & 5.2792 GeV \\
$M_{B_s}$ & 5.3693 GeV \\
$M_{K^0}$ & 0.498 GeV \\
$M_{\pi}$ & 0.140 GeV \\
$\Delta M_K$ & $5.301\times 10^{9}\,  \mbox{sec}^{-1}$ \\
\hline
$\omega$ & 0.045 \\
$\mbox{Re}A_0$ & $2.7\times 10^{-7}$ GeV \\
\hline \hline
$B_{7-9}^{1/2}$ & 1 \\
$\mu$ & $2$ GeV\\
\hline
\end{tabular}
\caption[]{ \it{Constants used in the numerical analysis.}}
\protect\label{tab:constants}
\end{center}
\end{table}
\renewcommand{\arraystretch}{1.3}
\item The statistical weight ${\cal W}_{i}$, suitably normalized to
the sum over all the events, is used to compute averages and errors
of the different quantities of interest. Since the probability
distributions are in general non-gaussian, we give the ``median'' and
the $68 \%$ confidence level intervals.  The median is defined in
such a way that half of the weighted {\it events} lies below its value.
The error range contains $68\%$ of the total weighted {\it events}.
Occasionally we will  also give  standard averages and errors,  or
ranges obtained by scanning the relevant parameters.
A meaningful definition of the scanning procedure requires the ``gaussian'' variables
to be extracted uniformly within some range. We choose a
range $\pm 1\sigma$ around the central values.
\end{enumerate}
\par We have divided the input parameters into two groups:
\begin{itemize}
\item[A)] in table~\ref{tab:constants},
 quantities for which the error is so small to give negligible effects in our
analysis are listed. In this table, we also give the value of the renormalization scale
$\mu$ at which Wilson coefficients and operator matrix elements are evaluated (this has
not been varied) and the values which have been taken for  $B_{7-9}^{1/2}$.
These matrix elements have never been computed on the lattice
and our choice is just a guess, biased by the VSA and justified by the fact
that the precise value of these $B$ parameters is not very important
to the estimate of $\epse$, see  fig.~\ref{fig:berto}.
\item[B)] in table~\ref{tab:variables},
 quantities which are extracted with gaussian (above the double horizontal line)
or flat (below the double line) distributions are listed. The value of $\Vub$ is our
average, and estimate of the error,  of the CLEO~\cite{CLEO} and LEP~\cite{LEP}
measurements.  The lattice results comes from  our compilation of several lattice
calculations.
In the table, only $B$ parameters of
left-left operators in the $HV$-scheme are listed. For these operators, the physical matrix elements can
be readily obtained  by multiplying physical quantities (such as
$f_K$ or $M_{K^0}$) times the $B$ parameters, since the quark masses never enter
their VSA expressions.
\end{itemize}
\renewcommand{\arraystretch}{1.0}
\begin{table}
\begin{center}
\begin{tabular}{|c|c|}
\hline
Parameter & Value and error \\ \hline
$V_{cb}$ & $0.0395 \pm 0.0017$ \\
$V_{ub}$ & $0.0037 \pm 0.0007$ \\
$\lambda$ & $0.2196 \pm 0.0023$ \\
$\alpha_s(M_z)$ & $0.119 \pm 0.003$ \\
$\ep$ & $(2.28 \pm 0.019) \times 10^{-3}$ \\
$\Delta M_{B_d}$ & $(0.472 \pm 0.016) \times 10^{12}\,  \mbox{sec}^{-1}$
\\
$\mtms$ & $165 \pm 5$ GeV \\ \hline \hline
$\mbms$ & $4.25 \pm 0.15$ GeV \\
$\mcms$ & $1.3 \pm 0.2$ GeV \\
$f_{B_d} \sqrt{\hat B_{B_d}}$ & $210 \pm 30$ MeV\\
$f^2_{B_d} \hat B_{B_d}/f^2_{B_s} \hat B_{B_s}$ & $1.14 \pm 0.06$ \\
$\Omega_{IB}$ & $0.25\pm 0.15$ \\
$\hat B_K$ & $0.87 \pm 0.13$ \\
\hline
\multicolumn{2}{|c|}{$B$ parameters in the $HV$-scheme at $\mu=2$ GeV} \\
\hline
$B_1^c$ & $0.075 \pm 0.075$ \\
$B_2^c$ & $0.075 \pm 0.075$ \\
$B_3$ & $3.5 \pm 2.5$ \\
$B_4$ & $3.5 \pm 2.5$ \\
$B_{5,6}$ & see eqs.~(\protect\ref{eq:q6hv}) and (\protect\ref{eq:vsandr}) \\
$B_{7,8}^{3/2}$ & see eq.~(\protect\ref{eq:q8hv}) \\
$B_9^{3/2}$ & $0.63 \pm 0.09$ \\
\hline
\end{tabular}
\caption[]{ \it{Variable parameters: average and errors, $\bar x \pm dx$.
The quantities above and below  the double horizontal line have been
extracted with gaussian and flat probability distributions,
respectively.   Matrix elements of left-left operators  are
expressed in terms of $B$ parameters and measurable quantities (such
$f_K$, $M_{K^0}$, etc.) only. In these cases,  it is equivalent to quote
the value of the matrix elements or the $B$ parameters. The latter are
listed in this table.}}
\protect\label{tab:variables}
\end{center}
\end{table}
\renewcommand{\arraystretch}{1.3}

Central values and errors of all the results presented in this study correspond to the ``median"
and the $68\%$ confidence-level region of the appropriate distributions.
In some cases, we also give ranges obtained by scanning all the input parameters.

\section{Constraints on the Unitarity Triangle}
\label{sec:triangle}
Analyses of the unitarity triangle and  NLO calculations of $\epse$ have been around
for several years~\cite{ciuc1}--\cite{buras2},\cite{reina1,mele,stocchi,re1}.
As discussed in the previous section, this analysis is based on  a comparison of theoretical
expressions for $\ep$, $\DMd$ and $\DMs$ with the
measurements/bounds and on the experimental determinations
of $\Vub$ and $\Vcb$.  
All relevant theoretical formulae can be found in refs.~\cite{ciuc1}--\cite{buras2}, 
\cite{silvestrini,reina1}.  Expressions, and numerical values, of the
Wilson coefficients appearing in the effective Hamiltonians, in all popular
renormalization schemes, can also be found in 
refs.~\cite{ciuc2,buras1,buras2},\cite{bur2}--\cite{noi}.
The values of the input parameters can be found in tables \ref{tab:constants} and
\ref{tab:variables}.

\begin{figure}[t]   
\begin{center}
\epsfxsize=\textwidth
\epsfysize=0.5\textheight
\leavevmode\epsffile{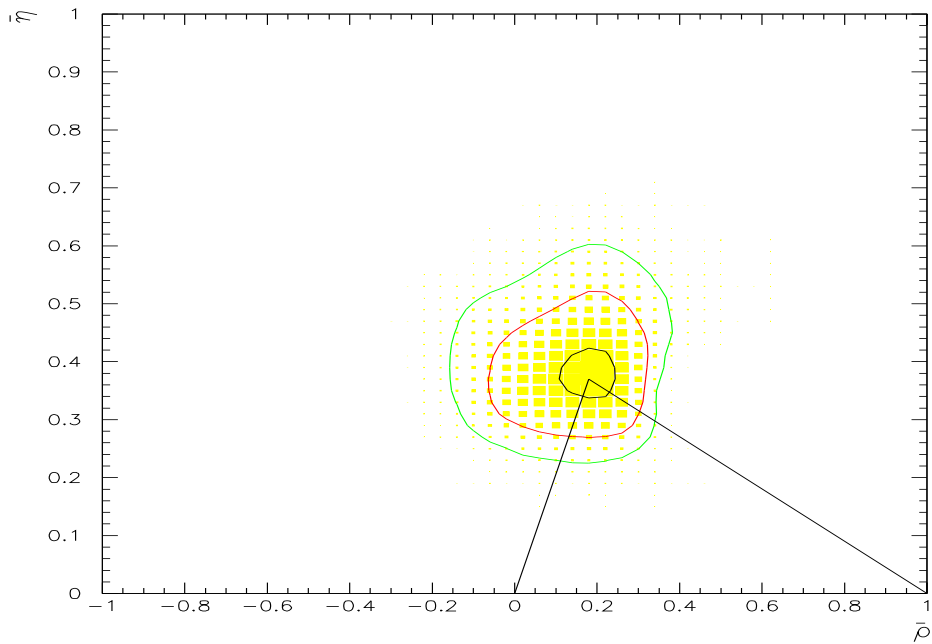}
\caption[]{\it{ Density plot in the $\bar\rho$--$\bar\eta$ plane. Contours
define regions containing  $5\%$, $68\%$ and $95\%$ of the generated events}}
\label{fig:triangle}
\end{center}
\end{figure}

The results of our analysis of the unitarity triangle are given in eq.~(\ref{eq:triangle}).
These values correspond to the density plot in the $\bar\rho$--$\bar\eta$ plane
shown in fig.~\ref{fig:triangle}.

Since the fit to the unitarity triangle is overconstrained, one can extract one (or more) input
variables, such as the renormalization group invariant $\hat B_K$ or $f_{B_d} \sqrt{\hat B_{B_d}}$,
together with the CKM parameters~\cite{mele,stocchi}.
By removing the  bounds on  one of the above  theoretical quantities,
we obtain respectively
\bea
\hat B_K= 0.90^{+0.26}_{-0.16} \, ,
\quad f_{B_d} \sqrt{\hat B_{B_d}}= 224^{+23}_{-25} \; \mbox{MeV}  \, .
\label{eq:fbfit}
\eea
These results are in very good agreement with  predictions  from lattice
QCD,  for example $\hat B_K= 0.89\pm 0.14$~\cite{lello99} (our average from a
compilation of lattice results for $\hat B_K$ is  given in table \ref{tab:variables}) and 
$f_{B_d} \sqrt{\hat B_{B_d}}= 210 \pm 20 \pm 20$ MeV~\cite{lubicz99} (for a recent review,
see also ref.~\cite{hashimoto99}). It is worth noting that lattice
predictions for these quantities existed long before the possibility of extracting
them from the analysis of the unitarity triangle and
have been very stable over the years~\footnote{Steve Sharpe estimated $\hat B_K= 0.84(3)(14)$
in 1996~\protect\cite{sharpe96}; a compilation by one of the authors of the present paper gave
$f_{B_d}  \sqrt{\hat B_{B_d}}= 220 \pm 40 $ MeV and $f_{B_d} \sqrt{\hat B_{B_d}}= 207 \pm 30$ MeV,
in 1995 and 1996 respectively~\protect\cite{marti96}.}.

\subsubsection*{Comparison with other analyses}

We now briefly compare our results with the two recent analyses of
refs.~\cite{mele} and \cite{stocchi}.

\begin{figure}[t]   
\begin{center}
\epsfxsize=\textwidth
\epsfysize=0.5\textheight
\leavevmode\epsffile{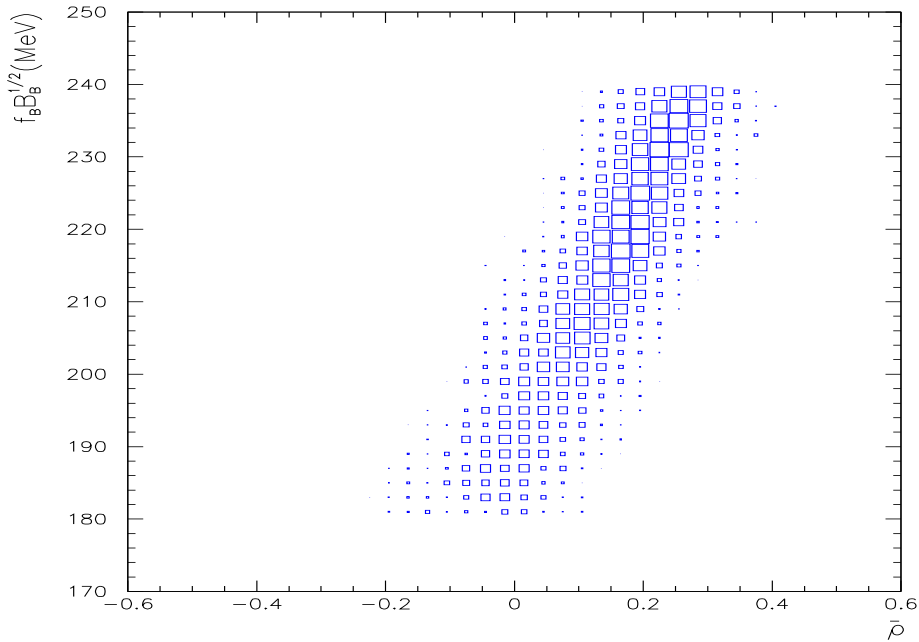}
\caption[]{\it{Density plot showing the correlation between $\bar\rho$ and $f_{B_d}
\sqrt{\hat B_{B_d}}$.}}
\label{fig:rhofb}
\end{center}
\end{figure}

In spite of several differences in the procedure followed in the analysis
of the data, our results in eq.~(\ref{eq:triangle}) are in very good agreement with
those of ref.~\cite{mele}, which quotes~\footnote{ In ref.~\cite{mele}, the
values of $\rho=0.160^{+0.094}_{-0.070}$ and
$\eta=0.381^{+0.061}_{-0.058}$ are given. We used $\overline\rho =
\rho (1- \lambda^2/2)$ and  $\overline\eta=\eta (1- \lambda^2/2)$, with the value
of $\lambda$ as given in table \protect\ref{tab:variables}.}
\be
\begin{array}{lll}
\overline\rho=0.156^{+0.092}_{-0.068}\,,& \overline\eta=0.372^{+0.060}_{-0.057}\, , &\\
\sin(2\alpha)=0.06^{+0.35}_{-0.42}\, , & \sin(2\beta)=0.75 \pm 0.090\, ,&
\gamma=(67^{+11}_{-12})^o\, .
\end{array}
\ee
Although our method is equivalent to that of
ref.~\cite{stocchi}, there are differences between their results,
\be
\begin{array}{lll}
\overline\rho=0.202^{+0.053}_{-0.059}\,, & \overline\eta=0.340\pm 0.035\, , &\\
\sin(2\alpha)=-0.26^{+0.29}_{-0.28}\, , & \sin(2\beta)=0.725^{+0.050}_{-0.060}\, , &
\gamma=(59.5^{+8.5}_{-7.5})^o \, ,
\end{array}
\ee
and those in eq.~(\ref{eq:triangle}). The main reason is the range chosen for
$f_{B_d} \sqrt{\hat B_{B_d}}$ (and to some extent for $\vert V_{cb}\vert$)~\footnote{ We thank A.~Stocchi for
pointing it out to us.}: in ref.~\cite{stocchi}
the  central value is similar to ours, and to the value used by ref.~\cite{mele}, but the
error is strongly asymmetric favouring  larger values of $f_{B_d} \sqrt{\hat B_{B_d}}$. This choice is
justified by  arguing that the existing unquenched results for  $f_{B_d}$
are larger than the  corresponding quenched ones. This is not a good argument, however,
because  the effect of quenching on the $B$ parameter, i.e. on the full matrix element
parametrized by  $f_{B_d} \sqrt{\hat B_{B_d}}$, is still unknown and could compensate the increase
of $f_{B_d}$.  Thus, for this quantity,
we have chosen the symmetric range in table \ref{tab:variables}.
Given the strong correlation between $\bar\rho$ and $f_{B_d} \sqrt{\hat B_{B_d}}$, as shown in
fig.~\ref{fig:rhofb}, this choice accounts for the bulk of the
differences in the results. We checked that, with a similar range for $f_{B_d} \sqrt{\hat B_{B_d}}$,
our results are much closer to those of ref.~\cite{stocchi}.
\par
Besides other minor differences in the range of the input parameters,
the various analyses also differ in the treatment of the errors. The authors
of ref.~\cite{stocchi} attempt to distinguish, for each input parameter,
statistical and systematic errors and assign to them different distributions.
In ref.~\cite{mele},  all the relevant input variables are included in the expression of
the $\chi^2$-function to be
minimized, implicitly assuming gaussian distributions for all these quantities.
In our case, we  have defined two classes of statistical- and systematic-dominated parameters (see table
\ref{tab:variables}), and use a single distribution, either gaussian or flat, for
each variable.
Finally, both refs.~\cite{mele} and \cite{stocchi} consider the Wilson coefficients relevant
to $K$ and $B$ meson mixing  as independent input parameters, assuming for them either gaussian
or flat distributions. This is not appropriate since the QCD corrections are known functions
of $\alpha_s$, $m_t$, etc. and not independent quantities.
In our analysis, for each event, we have computed the NLO QCD corrections,
thus taking correctly into account the correlations.

Apart from  the choice of the range of $f_{B_d} \sqrt{\hat B_{B_d}}$,
these further differences have small effect on the final results for the unitarity
triangle.
In addition, the fitted values of $\hat B_K$ and $f_{B_d} \sqrt{\hat B_{B_d}}$ in eq.~(\ref{eq:fbfit})
are well consistent with the results of ref.~\cite{stocchi}, where they found
$\hat B_K= 0.87^{+0.34}_{-0.20}$ and $f_{B_d} \sqrt{\hat B_{B_d}}= 223\pm 13$ MeV (Scenario I),
and with those of ref.~\cite{mele},
$\hat B_K= 0.80^{+0.27}_{-0.16}$ and $f_{B_d} \sqrt{\hat B_{B_d}}= 222^{+26}_{-11}$ MeV.

\section{Calculation of $\epse$}
\label{sec:epse}
In this section, we  summarize the main steps necessary to the calculation
of  $\epse$  and
give some details on the procedure followed to obtain the result quoted in
eq.~(\ref{eq:epseq6})--(\ref{eq:epse2}).   We also  make a critical comparison with previous
calculations of the Rome group~\cite{ciuc1}--\cite{ciuc3}, with the
recent estimates of ref.~\cite{silvestrini} and with results  obtained
in other approaches such as the $\chi$QM~\cite{bert1,bert2} and the $1/N$
expansion~\cite{paschoslast,pass}.

\par  Some general remarks are necessary before entering a more detailed
discussion. Given the large
numerical cancellations which may occur in the theoretical expression of $\epse$,
a solid prediction should avoid the ``Harlequin procedure". This procedure
consists in patching together $B_6$ from the $\chi$QM, $B_8$ from the
$1/N$ expansion, $\msms$ from the lattice, etc., or any other
combination/average of different  methods.  All these quantities
are indeed strongly correlated (for example $B_6$ and
$B_8$ in the $1/N$ expansion or  $B$ parameters and  quark masses
in the lattice approach) and should be consistently computed
within  each given theoretical framework.
 Unfortunately, as will clearly
appear from the discussion below, no one of the actual non-perturbative
methods is in the position to avoid completely the Harlequin procedure,
not even for the most important input  parameters only.
The second important issue is the consistency of the renormalization
procedure adopted in the perturbative calculation of the Wilson
coefficients and in the non-perturbative computation  of  the
operator matrix elements. This problem is particularly serious for the
$\chi$QM and the $1/N$ expansion, and will be discussed when comparing the
lattice approach to these methods. We will address, in particular, the
problem of the quadratic divergences appearing in the $1/N$ expansion.
This is an important issue, since the authors of
refs.~\cite{paschoslast,ham} find that these divergences
provide the enhancement necessary to explain the large values of
Re$A_0$ and of $\epse$.
\par Schematically, $\varepsilon^{\prime}$ can be cast in
the form \be \varepsilon^{\prime} = \frac{\exp (i\pi/4 )}{\sqrt{2}} \frac
{\omega}{\mbox{Re} A_{0}}\times \left[ \omega^{-1} \mbox{Im} A_{2} - (1-\Omega_{IB}) \mbox{Im}
A_{0} \right] \ee
where $\omega=\mbox{Re} A_{2}/\mbox{Re} A_{0}$  and $\mbox{Re} A_{0}$, given in table
\ref{tab:constants},  are taken from  experiments,  and
$\Omega_{IB}$ is a correcting factor, estimated in
refs.~\cite{don}--\cite{lu}, due
to isospin-breaking effects.  Using the operator product expansion,
the  $K \to \pi \pi$ amplitudes Im$A_{2}$ and Im$A_{0}$  are computed from the
matrix elements of the effective Hamiltonian, expressed in terms of
Wilson coefficients and renormalized operators
\be \langle\pi\pi\vert {\cal H}^{\Delta S=1}\vert K^0\rangle=
-\frac{G_{F}}{\sqrt{2}}
\sum_{i} C_{i}(\mu) \langle \pi\pi\vert Q_{i}(\mu) \vert
K^0\rangle \, \ee
where the sum is over a complete set of operators, which depend on
the renormalization scale $\mu$. In general, there are ten four-fermion operators
and  two  dimension-five  operators representing the chromo- and electro-magnetic dipole
interactions.
In the Standard Model, the contribution of these dimension-five operators
is usually neglected (possible SUSY effects can enhance the  contribution of the chromomagnetic
operator~\cite{murayama}) and, with  the scale $\mu=2$ GeV $> m_{c}$ at
which calculations are performed~\footnote{  $\mu = 2\div 3$ GeV $ > m_c $  is the typical
scale at which   matrix elements are computed in lattice QCD.}
only 9 out of the 10  four-fermion operators are  independent~\cite{bur3,noi}.
Wilson coefficients and   matrix elements of the  operators $Q_{i}(\mu)$,
appearing in the effective Hamiltonian, separately
depend on the choice of the renormalization scale and scheme. This dependence  cancels in
physical quantities, such as Im$A_{2}$ and Im$A_{0}$, up to higher-order corrections in the perturbative
expansion of
the Wilson coefficients. For this crucial cancellation to take  place, the non-perturbative method
used to compute hadronic matrix elements must allow a definition of the
renormalized operators consistent with the scheme used in the
calculation of the Wilson coefficients.
\par So far, lattice QCD is the only
non-perturbative approach in which both the scale and
scheme dependence can be consistently accounted for,
using either lattice perturbation theory or non-perturbative renormalization
techniques~\cite{npm,npm4f}. This is the main reason why we have followed
this approach over the years.

\begin{figure}[t]   
\begin{center}
\epsfxsize=\textwidth
\epsfysize=0.5\textheight
\leavevmode\epsffile{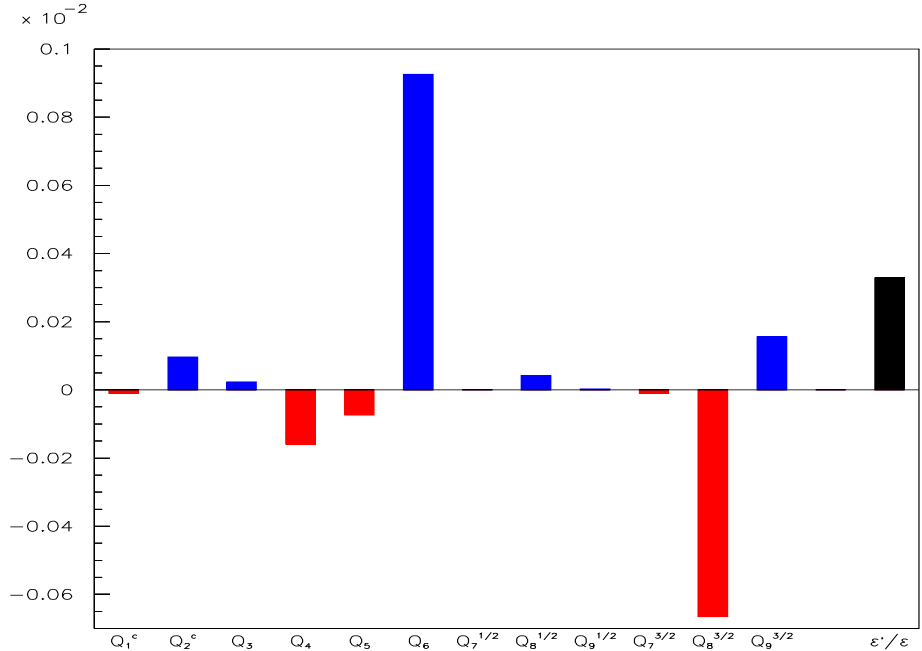}
\caption[]{\it{Individual contribution of the different operators to
$\epse$, using the value of $\langle Q_6\rangle_0$ in eq.~(\protect\ref{eq:q6hv}).
The corresponding value of $\epse$ is also shown.}}
\label{fig:berto}
\end{center}
\end{figure}
In fig.~\ref{fig:berto}, we display  the individual contributions
of each operator to the prediction of $\epse$ using matrix elements computed on the lattice,
as explained in the next section. Qualitatively, very similar  results are obtained also in other
approaches~\cite{bert2,silvestrini,paschoslast}.
There is a general consensus that the largest contributions to $\epse$ are those
coming from $Q_{6}$ and $Q_{8}$, with opposite sign, and sizeable
contributions may come from $Q_{4}$, $Q_{5}$  and $Q_{9}$
in the presence of large cancellations
between  $Q_{6}$ and $Q_{8}$, i.e. when the prediction for $\epse$
is of  ${\cal O}(10^{-4})$~\footnote{We recall the reader that $Q_8$ and $Q_9$ give large contributions
only to the I=2 amplitude.}. For this
reason the following discussion, and the comparison with other
calculations, will be focused on the determination, and errors, of the
matrix elements of the two most important operators. In the following, we adopt the following notation:
\be
\langle Q_j\rangle_{\{0,2\}}\equiv\langle\pi\pi\vert Q^{HV}_j(\mu=2\mbox{ GeV})\vert K^0\rangle_{I=\{0,2\}}
\ee
where the superscript denotes the $HV$ renormalization scheme
as defined in ref.~\cite{noi}~\footnote{
Incidentally, we note that the $HV$ scheme of ref.~\cite{bur2} is not the same as the $HV$
scheme of ref.~\cite{noi}.}.

\subsubsection*{Status of the calculation with matrix elements from lattice QCD}
The evaluation of physical $K \to \pi\pi$ matrix elements on the lattice relies
on the use of Chiral
Perturbation Theory ($\chi$PT): so far only $\langle \pi \vert Q_{i}(\mu) \vert
K \rangle$ and  $\langle
\pi(\vec p=0) \pi(\vec q=0) \vert Q_{i}(\mu) \vert K \rangle_{I=2}$
(with the two pions at rest) have been computed for a variety of operators. The
physical matrix elements  are then obtained by using  $\chi$PT at the lowest order.
This is a consequence of the difficulties in extracting physical multiparticle
amplitudes  in Euclidean space-time~\cite{mt}. Proposals to overcome this
problem have been presented, at the price of
introducing some model dependence in the lattice results~\cite{cfms}.
The use of $\chi$PT implies that  large systematic  errors may occur in the
presence of large corrections   from higher-order terms  in the chiral
expansion and/or from final-state interactions (FSI).
This problem is common to all approaches: if large higher-order terms
 in the chiral expansion are indeed present and
important,  any   method claiming to have these
systematic errors under control must be able  to reproduce
the FSI phases $\delta_0$ and $\delta_2$ of the physical amplitudes. The approaches of
ref.~\cite{bert1,bert2} and \cite{paschoslast}, however, give FSI
smaller than their physical values.
Regarding this issue, we note that the idea of improving the predictions of the hadronic
amplitudes using the experimental values of the FSI phases, with formulae such as
$\langle Q_i \rangle_I \to Re \langle Q_i \rangle_I/\cos \delta^{exp}_I$, is illusory.
If the FSI phases are not theoretically under control, one cannot tell whether the main uncertainty
comes from the real or the imaginary part of the computed amplitude, or from the absolute
value needed to compute $A_0$ and $A_2$.

\par\underline{$I=2$ matrix element of $Q_8$.} There exists a large set
of quenched calculations of $\langle Q_{8} \rangle_{2}$
performed with different formulations of the lattice fermion actions
(Staggered, Wilson, tree-level  improved, tadpole improved) and renormalization
techniques (perturbative, boosted perturbative, non-perturba\-tive), at
several values of the inverse lattice spacing $a^{-1}=2\div 3
$ GeV~\cite{giusti,npm4f},\cite{kgs}--\cite{lellolin}.
All these calculations, usually expressed in terms of
$B_{8}^{(3/2)}$,  give consistent results within $20 \%$ of uncertainty.
Among the results, we have taken the central value from the recent
calculation of ref.~\cite{giusti},
where the matrix elements  $\langle Q_{8} \rangle_{2}$ and
$\langle Q_{7} \rangle_{2}$ have been
computed directly without any reference to the quark masses, and inflated the
errors to account from the uncertainty due to the quenched
approximation (unquenched results are expected very soon) and the lack
of extrapolation to zero lattice spacing.
The values we use are
\bea
&&\langle Q_7\rangle_2 = 0.18 \pm 0.06 \, {\rm GeV}^{3} \, , \nn \\
&&\langle Q_8\rangle_2 = 0.62 \pm 0.12 \, {\rm GeV}^{3} \, .
\label{eq:q8hv}
\eea
The operator matrix elements computed without reference to quark
masses are given in physical units.
The reader who likes to work with $B$ parameters and quark masses may use
the formulae
\bea
\langle Q_7(\mu) \rangle_2 &=& \sqrt{2} f_\pi \left[
\frac{1}{3}\left( \frac{M_{K^0}^2}{\msms+\mdms}\right)^2-
\frac{M_{K^0}^2-M_\pi^2}{2} \right] B^{(3/2)}_7(\mu) \, ,\nn\\
\langle Q_{8}(\mu) \rangle_{2} &=& \sqrt{2} f_{\pi} \left[
\left( \frac{M_{K^{0}}^{2}}{\msms+\mdms}\right)^{2} -
\frac{ M_{K^{0}}^{2}  -M_{\pi}^{2} }{6} \right]  B^{(3/2)}_{8} ( \mu ) \, .
\eea
The values of the matrix elements
in eq.~(\ref{eq:q8hv}) correspond to $B_7^{(3/2)}=0.89\pm 0.30$ and
$B^{(3/2)}_8=0.93\pm 0.18$ for  a ``conventional'' mass  fixed to
$m_s^{\overline{MS}}+m_d^{\overline{MS}}=130$ MeV at $\mu=2$ GeV.
Anybody may rescale the value of the  $B$ parameters for these operators
according to her/his preferred value for  $\msms$.

\par\underline{Matrix element of $Q_6$.} For $\langle Q_{6} \rangle_0$
from the lattice, the situation
appears worse today than a few years ago when the calculations of
refs.~\cite{ciuc1}--\cite{ciuc3} were performed:
\begin{itemize}
\item[i)] until 1997, the only existing lattice
result, obtained with staggered fermions (SF) without NLO lattice perturbative
corrections,  was  $B_{6}=1.0 \pm 0.2$~\cite{kilcup}. This is the value
used in our previous analyses~\cite{ciuc1,ciuc2};
\item[ii)] with SF even more accurate results have been quoted  recently, namely
$B_{6}=0.67\pm 0.04\pm 0.05$ (quenched)  and  $B_{6}=0.76\pm 0.03\pm
0.05$ (with $n_{f}=2$)~\cite{k2};
\item[iii)] ${\cal O}(\alpha_{s})$
corrections, necessary to match lattice operators to  continuum
ones at the NLO, are so huge for $Q_6$ with SF (in the neighborhood of $-100 \%
$~\cite{stagpert}) as to make   all the above results unreliable.  Note,
however, that the corrections tend to diminish the value of $\langle
Q_6\rangle_0$;
\item[iv)] the latest lattice results for this matrix element, computed with domain-wall
fermions from $\langle \pi \vert Q_6\vert K\rangle$~\cite{newsoni},
are absolutely surprising: $\langle Q_6\rangle_0$ has the sign opposite to what expected
in the VSA,  and to what is found with the $\chi$QM and the $1/N$ expansion.
Moreover, the absolute value is so large as to give $\epse \sim - 120 \times 10^{-4}$.
Were this confirmed,  even the conservative statement by  Andrzej Buras~\cite{burask99}, namely
{\it ... that
certain features present in the Standard Model are confirmed by the experimental results. Indeed
the sign and the order of magnitude of $\epse$ predicted by the Standard Model turn out to agree with the
data...} would result too optimistic.
In order to reproduce the experimental number, $\epse \sim 20 \times 10^{-4}$, not only new
physics is required, but a large cancellation should also occur between the Standard Model
and  the new physics contributions.
Since this result has been obtained with domain-wall fermions, a lattice formulation
for which numerical studies began very recently, and no details  on
the renormalization and subtraction procedure have been given, we consider  it
premature to use the value of the matrix element of ref.~\cite{newsoni} in phenomenological
analyses. Hopefully, new lattice calculations will clarify this fundamental issue.
\item[v)]  estimates in the framework of the $1/N$ expansion
(where, however, one should always  to take into account the
correlation between the values of $B_{6}$ and $B^{(3/2)}_{8}$~\cite{1oN2})
and by the  $\chi$QM~\cite{bert2} give
$B_{6}=0.7\div 1.3$  at scale $\mu
=0.6 \div 1.0 $ GeV. One may argue~\cite{silvestrini,1oN2} that the scale
dependence of  $B_6$  above 1 GeV is rather weak and take the
range $B_{6}=0.7\div 1.3$ as valid also at $\mu=2$ GeV, which is the scale
at which we work.
Note, however, that the dependence of the matrix elements on the
renormalization scheme is rather strong and that, in these approaches,
the scheme   in which matrix elements are computed is unknown.
\end{itemize}
\begin{figure}[t]   
\begin{center}
\epsfxsize=\textwidth
\epsfysize=0.5\textheight
\leavevmode\epsffile{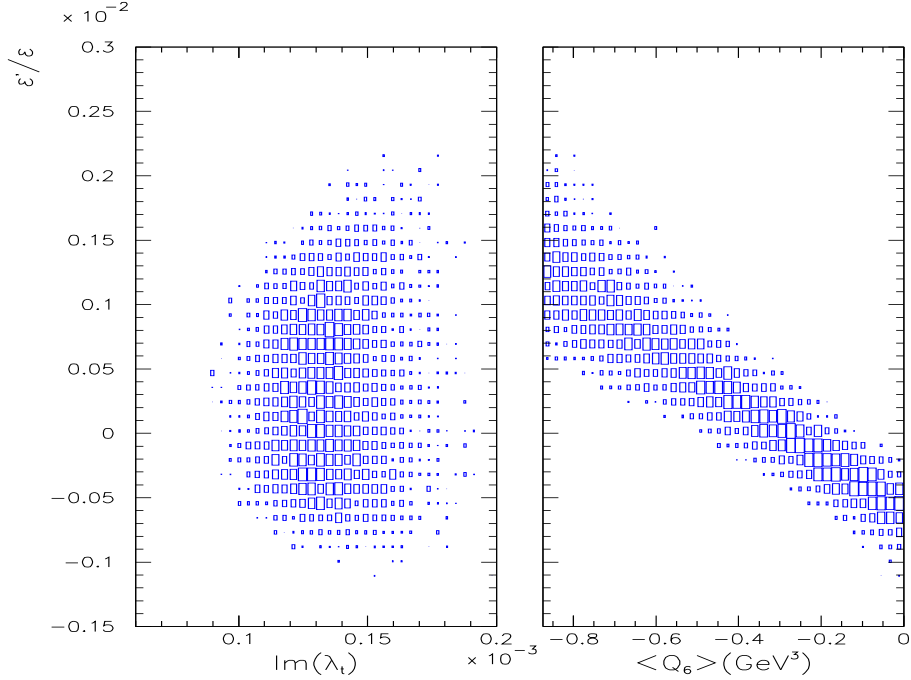}
\caption[]{\it{ Density plots showing the correlation of
Im $\lambda_t$ and $\langle Q_6 \rangle_0$ with $\epse$.}}
\label{fig:corre}
\end{center}
\end{figure}
Taking into account i)--v), we conclude that there is no computation
of $\langle Q_6 \rangle$ that can be reliably used in
phenomenological analyses.
For this reason, we present our result as in eq.~(\ref{eq:epseq6}).
In addition, biased by the VSA, we give results assuming
$B_{6}^{HV}=1.0\pm 1.0$ and $B_{6}^{HV}=1.35\pm 1.35$ (which corresponds
to $B_{6}^{NDR}=1.0\pm 1.0$), taking an error of $100\%$.
We compute the corresponding matrix element with the same
``conventional'' quark mass used for $\langle Q_{8} \rangle_{2}$, for
which an explicit  lattice calculation without reference to the quark
masses exists. In physical units, this choice corresponds to
\be
\langle Q_{6} \rangle_{0} = -0.4 \pm 0.4  \, {\rm GeV}^{3}\, ,
\label{eq:q6hv}
\ee
and
\be
\langle Q_{6} \rangle_{0} = -0.6 \pm 0.6  \, {\rm GeV}^{3}\, ,
\label{eq:vsandr}
\ee
and $\langle Q_{5} \rangle_{0}=1/3 \langle Q_6 \rangle_0$, in the two cases.
These expressions are obtained using the formulae
\bea
\langle Q_5(\mu) \rangle_0 &=&
 -\frac{4}{3} \left(\frac{M_{K^{0}}^{2}}{\msms+\mdms}\right)^{2}
 (f_{K}-f_{\pi}) B_{5}(\mu) \, ,  \nn\\
\langle Q_{6}(\mu) \rangle_{0} &=&
 -4 \left(\frac{M_{K^{0}}^{2}}{\msms+\mdms}\right)^{2}
(f_{K}-f_{\pi}) B_{6}(\mu)\,.
\eea
The values of $\epse$ obtained using the matrix elements in eqs.~(\ref{eq:q6hv}) and
(\ref{eq:vsandr}) are given in eqs.~(\ref{eq:epse1}) and (\ref{eq:epse2}) respectively.
The large difference between the two results is due
to the strong correlation between $\epse$ and $\langle Q_6 \rangle$, as shown in
fig.~\ref{fig:corre}. In the same figure we also show the correlation
between $\epse$ and $Im\lambda_t$, which is the parameter governing the
strength of CP violation in $K$-meson decays.
\begin{figure}[t]   
\begin{center}
\epsfxsize=\textwidth
\epsfysize=0.5\textheight
\leavevmode\epsffile{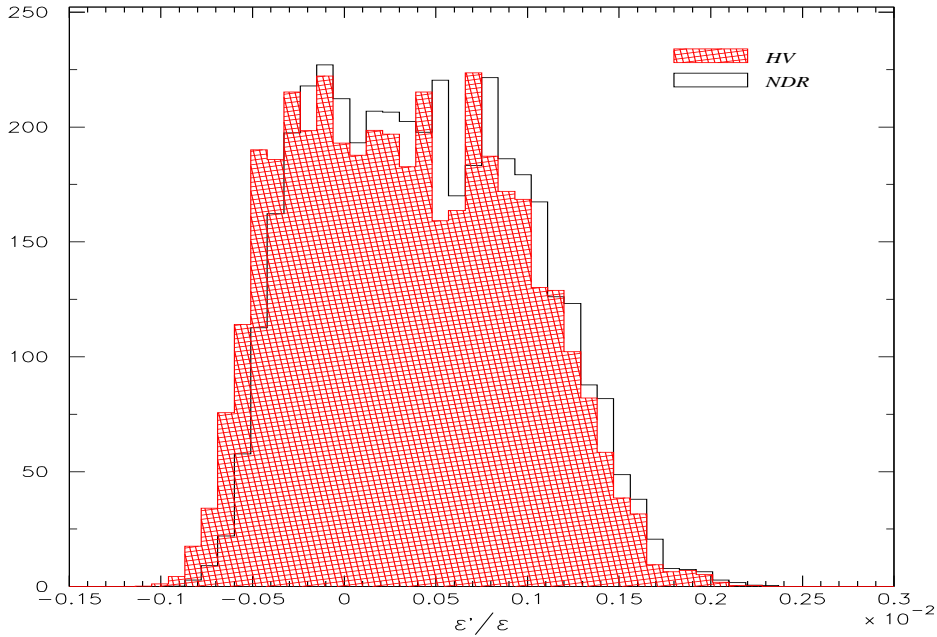}
\caption[]{\it{ Probability distribution of $\epse$ in the $HV$
and $NDR$ schemes. In changing schemes matrix elements and Wilson
coefficients have been consistently redefined at the NLO.}}
\label{fig:epsedist}
\end{center}
\end{figure}

In fig.~\ref{fig:epsedist}, we give two {\it event} distributions
of $\epse$, obtained using Wilson coefficients and hadronic matrix elements consistently
computed in $HV$ and $NDR$. Indeed, the calculation of the Wilson coefficients achieved in
refs.~\cite{ciuc2,buras1,buras2}, \cite{alta}--\cite{noi} allows a consistent
determination of the matrix elements of the renormalized operators at the NLO.
Notice that the two distributions in fig.~\ref{fig:epsedist} are quite similar, while
differences between matrix elements in different schemes can be
rather large: the parameter $B_6$ decreases, for instance, by
$\sim 30\%$ going from the $HV$ to the $NDR$ scheme.  This decrease is,
however, largely compensated by a readjustment of the corresponding Wilson
coefficients, as must happen in NLO calculations. The uncertainties due to higher order
perturbative corrections, given by the second error in eqs.~(\ref{eq:epse1}) and (\ref{eq:epse2}),
have been evaluated by modifying consistently Wilson coefficients and matrix elements
in the $HV$ and $NDR$ schemes.
In the two cases, using for example $\langle Q_6\rangle_0$ from eq.~(\ref{eq:q6hv}), we obtain
\bea
\epse&=&  (3.1^{+6.7}_{-6.3}  )\times 10^{-4} \quad HV \, , \nn \\
\epse &=&  (4.0^{+6.5}_{-6.1} )\times 10^{-4} \quad NDR\, ,
\eea
from which the result in eq.~(\ref{eq:epse1}) has been derived.

The results in eqs.~(\ref{eq:epse1}) and (\ref{eq:epse2})
are in very good agreement with previous estimates of the
Rome~\cite{ciuc1}--\cite{ciuc3} and Munich
group~\cite{buras1,buras2}. This agreement it is not surprising since the
two groups are using very similar inputs for the matrix elements and
the experimental parameters have only slightly changed in the last few
years.   The crucial problem, namely a  quantitative
determination of $\langle Q_6\rangle_0$, remains unfortunately  still
unsolved.
At present, we can only conclude that, with the central
value of $\langle Q_6\rangle_0$ taken from the VSA, even with a large error, it is difficult to
reproduce the experimental value of $\epse$.
On the other hand, by scanning various input parameters ($\asz$, $Im\lambda_t$, etc. and,
in the conventional approach, $B_{6}$ and $B_{8}^{(3/2)}$)
and in particular by  choosing them close to their extreme values,  it is
possible to obtain  $\epse$ up to $30 \times 10^{-4} $.
This gives  the impression of  a better agreement
(lesser disagreement) between the theoretical predictions and the data.
For completeness, we also give the interval of values of $\epse$
obtained by scanning, within one $\sigma$ the different parameters.
We obtain
\bea
-11 \times 10^{-4}  \le &\epse& \le  27 \times 10^{-4} \, , \nn\\
-10 \times 10^{-4}  \le &\epse& \le  30 \times 10^{-4} \, ,
\label{eq:scanning}
\eea
using $\langle Q_6\rangle_0$ from eqs.~(\ref{eq:q6hv}) and (\ref{eq:vsandr})
respectively.
Equation~(\ref{eq:scanning}) allows a direct comparison with several
calculations appeared in the literature
\be
\begin{array}{ll}
  0.2 \times 10^{-4}  \le \epse \le  22.0 \times 10^{-4} &
\mbox{Munich99-HV~\protect\cite{silvestrini}}, \\
  1.1 \times 10^{-4}  \le \epse \le  28.8 \times 10^{-4} &
\mbox{Munich99-NDR}, \\
  7.0  \times 10^{-4}  \le \epse \le  31 \times 10^{-4} &
\mbox{Trieste98~\protect\cite{bert2}}, \\
  1.5 \times 10^{-4}  \le \epse \le  31.6  \times 10^{-4} &
\mbox{Dortmund99~\protect\cite{paschoslast}}.
\end{array}
\ee
Note that our scanning results include a region of negative
$\epse$ on account of our choice of the error on $\langle Q_6\rangle_0$ being
larger than in other cases.

\par
In spite of the fact that the experimental world average  is compatible
with  the above ``scanned" ranges, we stress that, in order to get  a large
value of $\epse$,  a conspiracy of several inputs pushing $\epse$ in the same
direction is necessary. For central values of the parameters, the
predictions are, in general,  much lower than the experimental results.
For example,  the Rome-Munich and $1/N$  estimates are  typically
in the range $3$--$8 \times 10^{-4}$ and $8$--$10 \times 10^{-4}$,
respectively. For this reason,
barring the possibility  of new physics effects~\cite{murayama}, we
believe  that  an important message is arriving from the  experimental
results:
\par\noindent
{\it penguin contractions (or eye diagrams, not to be confused with penguin
operators~\cite{charming}), usually neglected within factorization,
give contributions to the matrix elements definitely larger than
their factorized values.}
\par
This implies that the ``effective'' $B$ parameters of the relevant  operators,
specifically  those relative to the matrix elements of $Q_{1}$ and $Q_{2}$ for
Re$A_{0}$ and of $Q_{6}$ for $\epse$ are much  larger than one. This
interpretation would provide a unique dynamical mechanism to explain   both
the  $\Delta I=1/2$ rule and   a large  value of $\epse$~\cite{ciuck99}.
Large contributions from penguin contractions  are  actually found
by calculations performed in the framework of the Chiral Quark
Model ($\chi$QM)~\cite{bert1,bert2} or the $1/N$
expansion~\cite{paschoslast,pass,1oN}--\cite{ham}.  It is very important that
these indications  find  quantitative confirmation in other approaches,
 for example in   lattice QCD calculations.  Note that  na\"{\i}ve
explanations of the large value of $\epse$, such as a very low value of
$\msms$, would leave the $\Delta I=1/2$ rule unexplained.

We have quantified the amount of enhancement required for the
matrix element
of $Q_{6}$ in order to explain the experimental value of $\epse$.
A fit of $\langle Q_{6} \rangle_{0}$ to  the world average
$Re(\epse)_{{WA}}$, using for the other parameters the standard values
given in tables \ref{tab:constants} and \ref{tab:variables} (in
particular by varying $\langle Q_{8} \rangle_{2}$ in the interval
given by eq.~(\ref{eq:q8hv}),
since for this operators penguin contractions are absent),
gives $\langle Q_{6} \rangle_{0}=-1.2^{+0.25}_{-0.21}\pm 0.15$ GeV$^{3}$, about
$2\div 3$  times  larger than the central values used in our analysis.
In  units which are more familiar to the reader, the value of  $\langle
Q_{6} \rangle_{0}$ required to fit the data
corresponds to  $B_{6}=2 \div 3$ for $m_s^{\overline{MS}}+m_d^{\overline{MS}}
=105 \div 130$ MeV (in the $HV$ scheme).

Before ending this  discussion, we wish  to  illustrate   the
correlation existing between  the $B$ parameters and the
quark masses  in lattice calculations.
\par On the lattice,  quark masses are often extracted
from the matrix elements of the (renormalized) axial current
($A_\mu$)   and pseudoscalar density ($P(\mu)$) (for simplicity we assume
degenerate quark masses)
\be   m(\mu) \equiv \frac{1}{2} \frac{\langle \alpha \vert \partial_\mu
A_\mu  \vert \beta \rangle}{\langle \alpha \vert  P(\mu) \vert \beta
\rangle} \, , \label{eq:awi} \ee
where $\alpha$ and $\beta$ are physical states (typically
$\alpha$ is the vacuum state and $\beta$ the one-pseudoscalar meson state)
and $m(\mu)$ and $P(\mu)$ are renormalized in the same scheme.
On the other hand, the $B$ parameters of  $Q_6$ and $Q_8$
are obtained (schematically) from the ratio of the following
matrix elements, evaluated  using  suitable ratios of
correlation functions~\footnote{ See for example ref.~\cite{npm4f}. We
omit the superscript $(3/2)$ in $B_8$ for simplicity.}:
\be B_{6,8}(\mu)  \propto \frac{\langle \pi \vert Q_{6,8}(\mu) \vert K
\rangle}{\langle \pi \vert P_\pi(\mu) \vert 0 \rangle\langle 0 \vert
P_K(\mu) \vert K \rangle} \, , \label{eq:bpar}\ee
where $P_\pi$ and $P_K$ are the   pseudoscalar densities with the flavour
content of the pion or  kaon, respectively.
Eqs.~(\ref{eq:awi}) and (\ref{eq:bpar}) demonstrate the strong correlation
existing between $B$ parameters and quark masses: large values of the
matrix elements of $P(\mu)$ correspond, at the same time, to small values
of $m(\mu)$ and  $B_{6,8}(\mu)$.
Physical amplitudes, instead, behave as
\be \langle Q_{6,8}\rangle = \mbox{const.} \times
\frac{B_{6,8}(\mu)}{m(\mu)^2}
\, , \label{eq:ratio}\ee
where  ``const." is a constant which may be expressed in terms of
measurable quantities (specifically $M_K$ and $f_K$) only.
From  eqs.~(\ref{eq:awi}) and (\ref{eq:bpar}),
we recognize that the dependence on $\langle P(\mu) \rangle$
cancels in the  ratio $B_{6,8}/m(\mu)^2$, appearing in
the physical matrix elements.
\par Previous lattice studies  preferred to work with
$B$ parameters because these are dimensionless quantities,  not
affected by the uncertainty due to the calibration of the lattice spacing.
This method can still be used, provided that quark masses and the $B$
parameters from the same simulation are  presented together
(alternatively one can give directly the ratio $B_{6,8}/m(\mu)^2$).
In ref.~\cite{giusti},  two possible definitions of dimensionless
``$B$ parameters", which can be directly related to physical matrix
elements without using the quark
masses have been proposed. In this analysis  we have used  the values of
$\langle Q_{7,8}\rangle_{I=2}$ computed  with one of these new definitions.

\subsubsection*{Comparison with the Munich group}
The original approach of the Munich group was to extract
the values of the relevant matrix elements from experimental
measurements~\cite{buras1,buras2}. This method guarantees the
consistency of the operator matrix elements  with the
corresponding Wilson coefficients.  In this approach, a  convenient choice
of the renormalization scale is $\mu=m_c$. In their analysis the
authors of ref.~\cite{silvestrini} used   the value $\mu=m_c=1.3$ GeV.
\par
Unfortunately, with the Munich method it is impossible to get
the two most important contributions, namely  those corresponding
to  $\langle Q_{6} \rangle_{0}$ and  $\langle Q_{8} \rangle_{2}$.
In this respect,
we completely agree with ref.~\cite{silvestrini} that one cannot
extract  $\langle Q_{6} \rangle_{0}$ from the experimental value of
Re$A_0$, unless further assumptions are made~\cite{ciuck99}.
For this reason, ``guided by the results presented above
and biased to some extent by the results from the large-$N$  approach  and
lattice calculations", the authors of ref.~\cite{silvestrini} have taken
$B_6 = 1.0 \pm 0.3$ and $B_8^{(3/2)}=0.8 \pm 0.2$, at $\mu=1.3$ GeV. These
values, if assumed to hold in the $HV$ regularization, are close to ours,
given the smooth behaviour of the $B$ parameters between $\mu=1.3$ and $2.0$
GeV.
The main differences in the evaluation
of  $\langle Q_{6} \rangle_{0}$ and  $\langle Q_{8} \rangle_{2}$
between ref.~\cite{silvestrini} and our calculation come from the value of
$\msms$  and from the scheme dependence that we now discuss.

In a complete NLO calculation, the scheme dependence of the matrix elements
is compensated by that of the Wilson coefficients up to NNLO terms, so that
physical quantities are independent of the renormalization procedure.
Lattice QCD allows a complete control of the definition of the renormalized
operators at the NLO: for example, $\langle
Q_8 \rangle_2$ has been computed with specific NLO definitions.
In ref.~\cite{silvestrini}, however, they kept fixed the values of the $B$
parameters when changing the renormalization scheme of the Wilson coefficients.
Although it is true that $\langle Q_6 \rangle_0$ is, at present, unknown, this procedure
introduces an unphysical scale and scheme dependence which should be avoided.
The use of the same $B$ parameters in two different schemes leads to
overestimate the error due to the scheme dependence.
We consider more appropriate to increase the error on the matrix element ($B$
parameter) in a given scheme and attribute the final uncertainty to our
ignorance on the matrix element, rather than to the choice of the
renormalization scheme.

\par For comparison, we also present the results obtained by taking the
main parameters close to those of ref.~\cite{silvestrini},
namely $B^{NDR}_6=1.0\pm0.3$, $B^{(3/2) NDR}_8=0.8 \pm 0.2$, $\msms=(110 \pm 20)$ MeV and the
condition $B_6 > B_8^{(3/2)}$. We find
\be
\epse =(7.2^{+3.6}_{-2.8}) \times 10^{-4}\, ,
\label{eq:comp}
\ee
a central value well consistent with the result of ref.~\cite{silvestrini}
in the $NDR$ scheme, given the remaining differences in the renormalization
scale and in the other matrix elements. However, we were not able to
reproduce the long positive tail in the $\epse$ distribution of ref.~\cite{silvestrini},
which produces an error more asymmetric than that in eq.~(\ref{eq:comp}).

\subsubsection*{$1/N$ expansion and $\chi$QM}
The $1/N$ expansion and the $\chi$QM are effective low energy theories
which describe the hadronic world.  To be specific, in the
framework of the $1/N$ expansion the starting point is given by
the chiral Lagrangian for pseudoscalar mesons expanded in powers
of masses and momenta. At the leading order in $1/N$ local  four-fermion
operators can be written as products of currents and densities,
which are expressed in terms of the fields and coupling of
the effective theory.  In higher orders, in order to compute
the relevant loop diagrams, a (hard) cutoff, $\Lambda_c$, must be
introduced. This cutoff
must be lower than $\sim 1$ GeV, since the effective theory only includes
pseudoscalar bosons and cannot account for vector mesons or heavier
excitations. The cutoff is usually identified with the scale at which
the short-distance Wilson coefficients must be evaluated. \par
Divergences appearing in factorizable contributions can be reabsorbed
in the renormalized coupling of the effective theory.
Non-factorizable corrections constitute the part
which should be matched to the short distance coefficients.
By using the  intermediate colour-singlet boson method, the authors
of refs.~\cite{paschoslast,bijnens} claim to be able to perform a
consistent matching, including the finite terms, of the matrix elements
of the operators in the effective theory to the corresponding Wilson
coefficients. It is precisely this point which, in our opinion, has never
been demonstrated in a convincing way.
\par If the  matching is ``consistent", then it should be possible
to show that in principle the cutoff dependence of the matrix elements computed in
the $1/N$ expansion cancels that of the Wilson coefficients,
at least at the order in $1/N$ at which they are working.
Moreover, if really finite terms are under control, it should be possible
to tell whether the coefficients should be taken in $HV$, $NDR$ or any
other renormalization scheme.
\par The fact that in higher orders even quadratic divergences appear,
with the result that the logarithmic divergences depend now on the
regularization, makes the matching even more problematic.
Theoretically, we cannot imagine any mechanism to cancel the cutoff
dependence of the physical amplitude in the presence of quadratic
divergences, which should, in our opinion, disappear in any reasonable
version of the effective theory.
Note that, in refs.~\cite{bert1,bert2}, the calculations are performed using
dimensional regularization in which quadratic divergences do not appear.
We suggest to repeat the calculation of the relevant matrix
elements with the $\chi$QM using a hard cutoff to show the stability of the
results with respect to the change of regularization scheme and verify the
possible presence of quadratic divergences. This would also provide an easier
comparison with the $1/N$ expansion.

It is also necessary to show
(and to our knowledge it has never been done) that the numerical
results for the matrix elements are stable with respect to
the choice of the ultraviolet cutoff.   This would also
clarify the issue of the routing of the momenta in divergent integrals.
For example, the matrix elements in the meson theory could be computed in some
lattice regularization.

\section{Conclusions}
\label{sec:conclusion}
In this paper, we have presented a combined analysis of the unitarity triangle
and $\epse$, using (whenever possible) matrix elements from lattice QCD, which is
theoretically well suited for NLO calculations.
We stress that, given the correlations among different non-perturbative parameters,
calculations of $\epse$ should use matrix elements consistently computed within a given theoretical approach,
at least for the main contributions.
At present, however, there is no reliable calculation of $\langle Q_6\rangle_0$ on the lattice.
For this reason, our main result is the one given in eq.~(\ref{eq:epseq6}), in which the
matrix element is left as a free parameter.
In addition, we give the results in eqs.~(\ref{eq:epse1}) and (\ref{eq:epse2}), by taking the
central value of $\langle Q_6\rangle_0$ from the VSA in
different renormalization schemes with a relative error of $100\%$.
Our results show that, even with such a large
error, it is difficult, although not impossible, to reproduce the experimental value of $\epse$.
To this end, a conspiracy of several input parameters, pushing in the same direction, is necessary.
We rather think that the important message arriving from the experimental
results is that penguin contractions (eye diagrams)
give  contributions which make the matrix element of $Q_6$ definitely larger than
what expected on the basis of the VSA. This interpretation provides a unique dynamical mechanism to
account for both the  $\Delta I=1/2$ rule and a large value of $\epse$
within the Standard Model, whereas other arguments, as those based on a low value of $m_s^{\overline{MS}}$,
would leave the $\Delta I=1/2$ rule unexplained.
Concerning the strange quark mass, we stress that $m_s^{\overline{MS}}$ is irrelevant for the calculation of
the operator matrix elements on the lattice.

In the long run, lattice QCD is the only non-perturbative method able to produce quantitative
results at the NLO accuracy. In the present situation, however, other approaches, such as the $\chi$QM or the
$1/N$ expansion, which cannot control the proper definition of the renormalized operators, may prove useful
to understand the underlying dynamics.
We hope that the issue of the computation of  $\langle Q_{1,2}\rangle$ and
$\langle Q_6\rangle_0$ on the lattice will be clarified soon, providing reliable theoretical
estimates for both the $\Delta I=1/2$ rule and $\epse$.

\end{document}